\documentclass[useAMS,usegraphicx,usenatbib]{mn2e}
\usepackage{aas_macros}     

\usepackage{fixltx2e}[1999/12/01]

\newcommand{\Mass}{ \ensuremath{ M_{\odot}} }

\newcommand{\hMpc}{ \ensuremath{\rm h^{-1} Mpc} }

\title[The Birth and Growth of Neutralino Haloes]
{The Birth and Growth of Neutralino Haloes}
\author[Angulo  et al.]{
\parbox[h]{160mm}{
R. E. Angulo\thanks{reangulo@mpa-garching.mpg.de} \& S. D. M. White
}\vspace{6pt}\\
Max Planck Intitute fur Astrophysik, D-85741 Garching, Germany.
\vspace*{-0.5cm}}

\begin{document}
\date{\today}
\pagerange{\pageref{firstpage}--\pageref{lastpage}} \pubyear{2009}
\maketitle
\label{firstpage}

\begin{abstract}
We use the Extended-Press-Schechter (EPS) formalism to study halo assembly
histories in a standard $\Lambda$CDM cosmology. A large ensemble of Monte Carlo
random walks provides the {\it entire} halo membership histories of a
representative set of dark matter particles, which we assume to be neutralinos.
The first generation halos of most particles do not have a mass similar to the
free-streaming cut-off $M_{f.s.}$ of the neutralino power spectrum, nor do they
form at high redshift. Median values are $M_1 = 10^5$ to $10^7M_{f.s.}$ and
$z_1 = 13$ to $8$ depending on the form of the collapse barrier assumed in the
EPS model. For almost a third of all particles the first generation halo has
$M_1>10^9M_{f.s.}$. At redshifts beyond 20, most neutralinos are not yet part
of any halo but are still diffuse. These numbers apply with little modification
to the neutralinos which are today part of halos similar to that of the Milky
Way. Up to $10\%$ of the particles in such halos were never part of a smaller
object; the typical particle has undergone $\sim 5$ ``accretion events'' where
the halo it was part of falls into a more massive object. Available N-body
simulations agree well with the EPS predictions for an ``ellipsoidal'' collapse
barrier, so these may provide a reliable extension of simulation results to
smaller scales. The late formation times and large masses of the first
generation halos of most neutralinos imply that they will be disrupted with
high efficiency during halo assembly.
\end{abstract}


\section{Introduction}

The existence of ``Cold Dark Matter" (CDM) is one of the pillars of our current
understanding of the origin and evolution of structure in our Universe.  Many
independent astrophysical observations have not only provided evidence of its
existence but also indications of its properties; the CDM particle should be
non-baryonic, collisionless, neutral and with small initial velocities. In
spite of the arguments supporting such a particle, no member of the standard
model of particle physics can fit the requirements. Fortunately, this problem
is ``solved" in supersymetric extensions where many candidate particles for CDM
emerge. In particular, the lightest neutralino, which should have a mass of
about $100$ GeV, is the currently favoured choice, as it is both weakly
interacting and stable \citep[see][for a review of current candidates]{Bertone2005}.

Some properties of the neutralino, such as its mass, have a direct impact on
the formation and evolution of structure in the Universe. The finite
temperature at which these particles decouple from the radiation field before
recombination imprints features in the primordial power spectrum of
fluctuations at very small scales. In particular, the free streaming of
neutralinos suppresses perturbations below $\sim 0.7\,{\rm pc}$ and, as
a consequence, the smallest objects that can gravitationally collapse have a
mass $M_{f.s.} \sim 10^{-8}\Mass$. Naturally, this will significantly affect
the properties of the first objects in the Universe, which must have masses
comparable to, or larger than, $M_{f.s.}$. The free streaming of
neutralinos may also modify the way in which much larger haloes grow. For
example, the amount of mass that a halo accretes in a diffuse form depends on
how much mass in the Universe is in collapsed objects. The objective of this
paper is to study these and other aspects of structure formation in a
$\Lambda$-neutralino cosmology. 

The most accurate way to study the highly nonlinear dynamics involved in the
formation and evolution of haloes is via N-body simulations.  However,
resolving galactic haloes and objects with $M \sim M_{f.s.}$ simultaneously,
poses an extremely hard problem that is currently impossible to solve.  For
example, a direct N-body simulation of the Milky Way's halo would require at
least $10^{23}$ particles, almost $14$ orders of magnitude larger than the most
sophisticated simulations performed so far. Extrapolating Moore's law, such
calculation may become possible after the year $2050$.

So far, several different approaches have been used in the literature to study
the implications of neutralinos for early structure formation.  One consists in
simulating scale-free cosmologies. There, the initial power spectrum is assumed
to be a power law, with index similar to that of CDM on very small scales.
Although the range of scales is no larger than in a standard CDM simulation, it
superficially appears possible to study the formation of very small haloes,
with masses similar to the free streaming cut-off mass \citep[e.g][]{Widrow09}.
The weakness of such simulations is that the results are not properly coupled
to the evolution of longer wavelength perturbation; as we shall see below,
these are actually very important in a realistic CDM cosmology. Another tactic
is to resimulate extremely low density regions at very high resolution in such
way that the Lagrangian region of the simulation is confined to a small zone
and the mass of each simulation particle can be very small
\citep[e.g][]{Diemand05}. This approach is intrinsically limited to
unrepresentative regions of the very high-redshift Universe. Yet another tactic
is to carry out a set of nested zoomed simulations \citep{GaoWJF2005} although this
technique has yet to be extended all the way to the free-streaming mass.

In this paper we adopt a less accurate but self-consistent strategy based on
excursion set theory \citep{PressSchechter1974, Bond1991, Bower1991,
LaceyCole1993}. This formalism has been extremely successful in reproducing
many aspects of dark matter halo formation, as well as in fitting halo mass
distributions and clustering. It provides a realistic model for halo growth
over its entire history.  Although the theory has not been tested on very small
scales, it is currently the only way to compute the {\it full} mass assembly
history of present-day haloes, i.e.  starting from the free-streaming mass, the
very bottom of the CDM hierarchy, and following growth up to cluster scales,
the largest collapsed objects in the Universe.  

This paper is organized as follows. First, we illustrate how the free-streaming of
neutralinos modifies the the primordial density field. In \S3 we provide the
theoretical background and the tools necessary for our analysis.  We discuss our
first results in \S4, where we investigate the typical redshift at which dark
matter particles become part of a halo for the first time, as well as the
mass of those haloes. We then move to \S5 where we look at the assembly history
of a Milky-Way sized halo. We present some final remarks in \S6.

Note that thoughout this paper we will use the following cosmological model:
matter density parameter,  $\Omega_M= 0.25$, vacuum energy density parameter,
$\Omega_\Lambda = 0.75$, normalization of density fluctuations, expressed in
terms of the extrapolated linear amplitude of density fluctuations in spheres
of radius $8\,\hMpc$ at the present-day, $\sigma_8=0.9$, primordial spectral
index  $n_s=1$ and Hubble constant,  $H_0 = 73\,{\rm km\,s^{-1}\,Mpc^{-1}}$. We
will also, in general, adopt a standard neutralino with mass $100$ GeV.

\section{Neutralino CDM Power spectrum}

\begin{figure} 
\includegraphics[width=8.5cm]{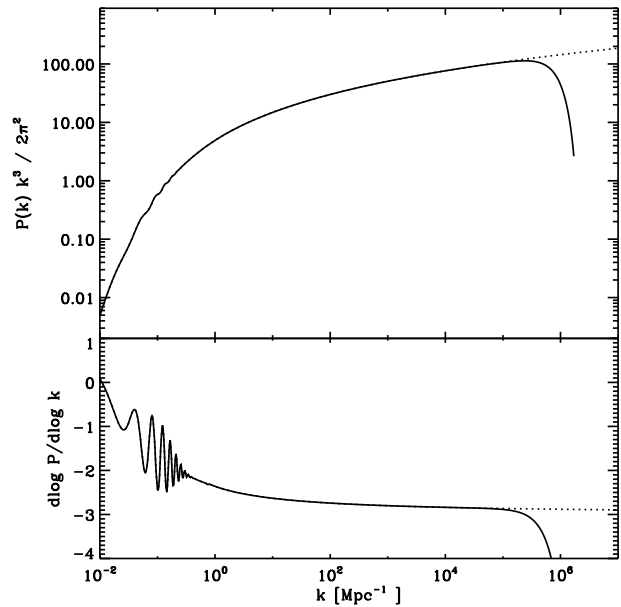} 
\caption{ {\it Top :} The dimensionless power spectrum of the dark matter
density field at $z=0$ predicted by linear perturbation theory.  {\it Bottom :}
The logarithmic derivative of the power spectrum, i.e. the local power law
index of the power spectrum.  The solid line shows the power spectrum assuming
that the dark matter particles are neutralinos of mass $100$ GeV whilst the
dashed lines effectively assume an infinite mass for the dark matter particle.
\label{fig:pk}} \end{figure}

\begin{figure} 
\includegraphics[width=8.5cm]{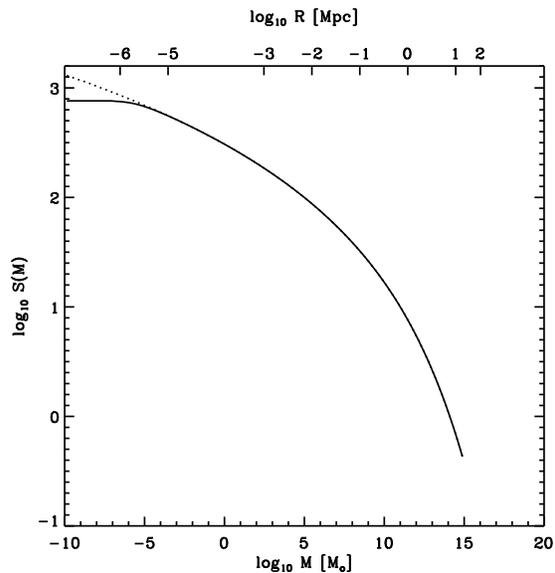} 
\caption{The logarithm of the $z=0$ mass variance $S(M)$ smoothed with a real space
top-hat filter as a function of the smoothing radius (mass) normalised to
$\sigma_8=0.9$.  The solid lines lines take into account the cut-off in the
power spectrum for $100$ GeV neutralinos.
\label{fig:sigma}} \end{figure}

In the classical $\Lambda$CDM model, the density fluctuations increase
monotonically with wavenumber. However, this behaviour is modified by
neutralino streaming. The mass of the neutralino sets the temperature at which
the population becomes nonrelativistic.  This moment sets the primordial
velocity dispersion of neutralinos and characterizes the scale below which
perturbations are suppressed.  Roughly speaking, any perturbation smaller than
the mean free path of neutralinos over a Hubble time will be dissipated,
producing a cut-off in the primordial power spectrum. 

We can see this effect quantitatively in Fig.\ref{fig:pk}. This plot shows the
neutralino-CDM power spectrum predicted by linear perturbation theory.  The
corresponding variance of the linear density field, extrapolated to $z=0$, in
spheres of different radius (and hence enclosed mass) is shown in
Fig.\ref{fig:sigma}. We have computed fluctuations on large scales using the
Boltzmann code {\tt CAMB} \citep{Lewis2000} whilst on very small scales, the
power spectrum is that predicted by the approach of \cite{Green04} which
includes effects from the free streaming of neutralinos . 

In both figures, solid lines assume a neutralino of mass $100$ GeV and a
temperature at kinetic decoupling of $33$ MeV. As previously discussed, free
streaming suppresses the power on very small scales ($R_{f.s.} \sim 0.5\,h^{-1}
{\rm pc}$) generating an exponential cut-off.  The counterpart of the lack of
power on small scales is a flattening of the variance for very small smoothing
radii.  Thus there is a finite maximum variance of the field which, for our
cosmological model, is equal to $\sim 760$.  For comparison we have also
included the prediction for a very massive CDM particle which is displayed as
dotted lines. In this case there is no suppression of fluctuations and
the variance diverges logarithmically as the slope of the power spectrum
approaches $-3$. 

The features introduced by the neutralino have immediate implications for
structure formation. For instance, the smallest possible dark matter objects in
the Universe have a mass $\sim 10^{-8}\Mass$ ($1\%$ of the mass of the Earth)
and the most abundant haloes are predicted by excursion set theory to have
masses typically of order $10^{-6}\Mass$.

The existence of a minimum halo mass in the Universe, does not, however, imply
that most of the dark matter was once part of such an object. Indeed, as we
will show in subsequent sections, only a small fraction of neutralinos were
ever part of such haloes.  Before we investigate this and related issues in
more detail, the following section gives a brief overview of the techniques
that will make our analysis possible.   

\section{Methodology}

We start this section by reviewing excursion set theory and the various
collapse models that are used as part of it. We also discuss our practical
implementation of the formalism. Finally, we provide a description of the
N-body simulation with which we will test some of our results.

\subsection{The excursion set approach to the formation and evolution of
haloes}

Dark matter haloes are highly nonlinear objects. In spite of their complexity,
many of their statistical properties can be described surprisingly
well by the simple analytical arguments of excursion set theory. In the
following we will give a brief summary of the main ideas behind the model, a
comprehensive review is given by \cite{Zentner2007}.    

The classic implementation of excursion set theory \citep{Bond1991,
Bower1991, LaceyCole1993} starts by considering a random particle and then
smoothing the density contrast field around it on progressively smaller scales
until its smoothed overdensity crosses a threshold for collapse. The particle
then is assumed to belong to a halo of Lagrangian mass equal to that enclosed
in the smoothing window at threshold. 

In practice, the smoothed density contrast is generated by convolving the
density contrast field with a smoothing window, $W$, 

\begin{equation} 
\delta_R(\vec{x}) = \int \delta(\vec{x'}) W(\vec{x}-\vec{x'}; R) {\rm d}^3 x'.
\end{equation}

On the same scale the total variance of the field, $S(R)$, is:

\begin{equation} 
S(R) \equiv \langle \delta_R^2(x) \rangle 
 	= \frac{1}{2\pi^2} \int {\rm d}k k^2 P(k) \widetilde{W}^2(k; R),
\end{equation}

\noindent where $P(k)$ is the dark matter power spectrum discussed in the
previous section.  An extremely interesting filter is a top-hat in Fourier
space, where $\widetilde{W}(k; R)= 1$ for all points with $k \le R^{-1}$ and
$\widetilde{W}(k; R)= 0$ otherwise. In this case the smoothed density contrast
executes a Markov random walk in the $S-\delta$ plane as $R$ decreases, since 
increasing the window adds wavemodes that are independent of those previously included.

Once we have computed the trajectory $\delta(S)$ for each particle, the next
step is to associate the particle to a collapsed object.  The simplest
criterion for collapse is that a dark matter particle is considered part of a
halo with mass $M$, such that the density contrast smoothed on the scale $R(M)$
first exceeds some critical value.  The relation between $R$ and $M$ is given
by the volume enclosed in the smoothing window. 

The critical overdensity for collapse at a given redshift may be taken as constant
or as function of the variance; these are known as constant and moving barriers
respectively.  The first is motivated by the spherical collapse model,
which predicts that at the moment of virialisation the linearly interpolated
density contrast is $\delta_{sc} \sim 1.686$. On the other hand, the moving
barriers approximately account for the fact that, in order to collapse, a small
perturbation must have a higher density contrast compared with a larger
perturbation because it is typically more asymmetrical. These barriers are
motivated by ellipsoidal collapse models and they are usually taken to have the following form:

\begin{equation} 
\delta_c(S,z) = \sqrt{q}\delta_{sc} \left[1+\beta \left(\frac{S}{q \delta_{sc}^2} \right)^\gamma \right],
\end{equation}

\noindent  where, by using the expectation value for the shape of dark matter
haloes as a function of scale, a single condition (depending only on the variance of
the field) can be applied to all trajectories at a given redshift.

By analysing the criteria for collapse for an ensemble of walks as a function
of redshift, the excursion set formalism predicts the probability
distribution of halo assembly histories for an ensemble of DM particles.

Despite the simplicity of the argument, the excursion set approach has been
extremely successful in reproducing many properties of DM haloes as determined
from N-body simulations. In particular the clustering and number density of
haloes are well reproduced, as are the mass function of their progenitors.
Nevertheless, the approach fails to reproduce all results from simulations, for example,
the dependence of clustering on halo properties other than mass \citep{Gao2005,
Wechsler2006, GaoWhite2007, Wetzel2007, Angulo2008b}. This particular disagreement arises
because the Markov nature of the overdensity trajectories implies that there can
be no correlation between the assembly history of a halo and its large-scale
environment \citep{White1996}. Such a dependence is, however, found in simulations.

\subsection{Monte Carlo Simulations}

We have implemented the ideas described in the previous subsection as follows.
First we compute the smoothing radii for which we will generate each random
walk. These have a variable spacing in mass given by the condition $M_{i} = 0.9
M_{i-1}$. This choice ensures that we can easily resolve all the events where
the halo associated with a random walk increases its mass by a factor of two,
allowing us to resolve all infall events and major mergers.  The probability
that at a given radius $R_{i}$ the smoothed field has a value between $\delta$
and $\delta+{\rm d}\delta$ is then given by $\exp[(\delta-\delta_{\rm
i-1})^2/S(R_i)] {\rm d}\delta$. 

This approach, combined with the maximum finite variance of the field (c.f.
\S2) implies that only $\sim 2000$ random numbers are necessary to follow the
complete assembly history of the haloes associated with a particular dark
matter particle over the $25$ orders of magnitude which separate from the
smallest halo it could possibly reside in from a $M_*$ halo today. 

At each step of the random walks we check whether a collapse criterion is
fulfilled or not.  We considered both the constant spherical collapse barrier
and a square root barrier for which $(q,\beta,\gamma) = (0.5,0.55,0.5)$ in
Eq.~3.  Note that the latter set of parameters ensures that barriers at
different redshift will never intersect each other, as can happen, for
instance, for the ellipsoidal barrier that best reproduces mass functions in
N-body simulations \citep{ShethMoTormen2001,Mahmood05,Moreno08}.

We have used two different starting points for the density and variance.  First
we set $(S,\delta)_{i=0} = (0,0)$ so the trajectories represent a random set of
CDM particles. These walks allow us to investigate general properties of dark
matter haloes which will be presented in \S5. The second starting point is
$(S,\delta)_{i=0} = (4.96, 1.686)$ and $(S,\delta)_{i=0} = (4.96, 2.36)$  for
the spherical collapse and square root barrier respectively, which generates
trajectories that represent CDM particles that end up in a Milky Way sized halo
today. In this way we can look into the properties of the mass assembly of such
haloes. We will analyse these in \S6. 

We repeat our algorithm $10^5$ times to generate a large ensemble of random
walks. This allow us to represent the probability distribution of halo mass
assembly histories for neutralinos in a $\Lambda$CDM Universe.

\subsection{The hMS simulation}

The smallest haloes that can be resolved in current N-body simulations are many
orders of magnitude more massive than the smallest haloes expected in a
neutralino-CDM Universe. Nevertheless, by imposing on our random walks an
artificial minimum halo mass that matches the resolution of an N-body
simulation, we can quantitatively assess the performance of the excursion set
approach.

The N-body simulation we have chosen to compare our results with is the hMS.
This calculation used $900^3$ particles, each mass $1.3 \times 10^{8}\Mass$, to
solve the gravitational dynamics of the dark matter distribution in a periodic
box of side $137\,{\rm Mpc}$.  The set of cosmological parameters is exactly
the same as those used in the Millennium Simulation \citep{Springel2005b}.

Every $\sim100$ Gyr we have identified haloes using a FoF algorithm
\citep{Davis1985} as well as substructures within them using the {\tt SUBFIND}
procedure described in \cite{Springel2001}. Furthermore, we have found the
progenitor and descendants of every halo by following its 10\% most bound
particles across different snapshots, as described in \cite{Angulo09}. The
combination of high-resolution in mass and a considerable volume allows us to
sample the evolution of dark matter particles over a wide range of halo mass. 
 
\section{The first objects in the Universe}

In this section we present statistics for the first haloes occupied by typical
DM particles based on our ensemble of random walks. In particular, we will show
the total mass fraction in haloes of {\it any} mass, and the distribution of
mass and redshift at which typical neutralinos become part of a halo for the
first time. 

\subsection{The mass in collapsed objects}

\begin{figure} 
\includegraphics[width=8.5cm]{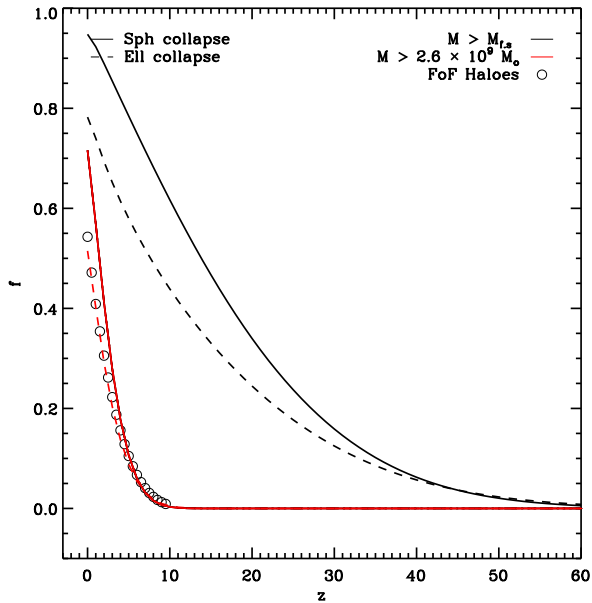} 
\caption{The fraction of mass in collapsed objects of any mass (black curves)
and above a mass of $2.6\times10^{9}\Mass$ (red curves) as a function of
redshift.  The results displayed as solid lines are computed using a constant
barrier while the dashed lines are computed using a square root barrier.
Measurements from an N-body simulation are also displayed as black circles. Note
that the black lines assume no cut other than that imposed by the nature
of the neutralino.
\label{fig:massinh}} \end{figure}

Fig.\ref{fig:massinh} shows as a function of redshift, the fraction of random
walks that have ever crossed the threshold for collapse, as well as the
fraction that cross for a value of the smoothing mass greater than $2.6 \times
10^{9}\Mass$. These curves correspond to the fraction of all cosmic matter in
halos of {\it any} mass, and in haloes above $2.6 \times 10^{9} \Mass$. 

A common misconception is that in CDM models every DM particle belongs to a
halo of some mass. Fig.\ref{fig:massinh} clearly contradicts this.  The
fluctuation cut-off induced by neutralino streaming implies that some
trajectories never cross the critical threshold for collapse. As we follow
trajectories down to smaller smoothing masses, the fraction that have crossed
the threshold asymptotes to values smaller than unity. The asymptotic value
depends on the shape of the barrier, the fraction of all matter predicted to be
in diffuse form at $z=0$, i.e. to be part of no clump, is $5\%$ and $22\%$ for
constant and square root barriers, respectively. 

This effect is stronger at at high redshift where a much lower fraction of DM
is part of haloes. Less than half the mass is in any halo beyond redshifts
$7.8$ and $13.8$, for the ellipsoidal and spherical models, respectively.  By
$z \sim 34$ the fraction of diffuse mass has reached $90\%$ in both models.
Finally, beyond $z\sim 60$ less than $1\%$ of the mass is in gravitationally
bound structures.

Black circles in Fig.\ref{fig:massinh} show the fraction of DM particles
associated to FoF haloes (identified with at least $20$ particles) in the hMS
simulation. This can be compared directly with the red lines which indicate the
fraction of random walks in haloes above the corresponding mass limit $M >
2.6\times 10^9\Mass$.  For the square root barrier  $49.4\%$ of the random
walks are associated with such haloes at $z=0$, this is very close to the
fraction of the particles found in FoF haloes ($54.3\%$).  A larger difference
exists with the spherical model which overestimates the mass fraction in haloes
at redshift zero by roughly $30\%$.  In general, however, the agreement with
the two models is quite good.

\subsection{The mass and redshift distribution of first objects}	

\begin{figure} 
\includegraphics[width=8.5cm]{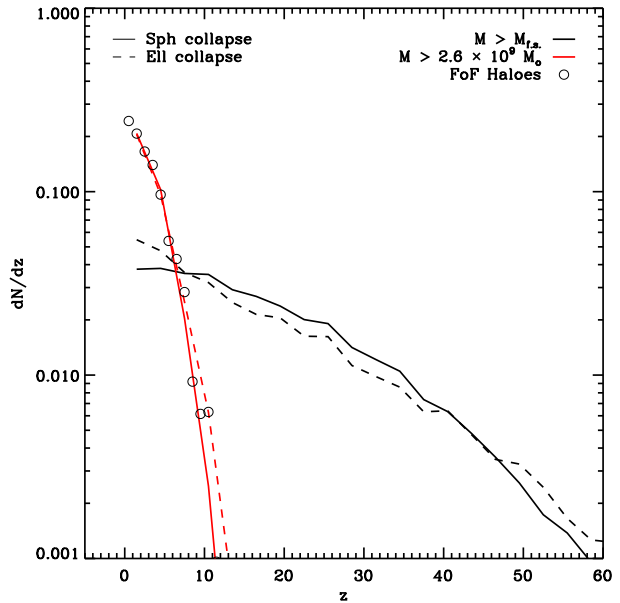} 
\caption{ The distribution of redshifts at which DM particles are accreted onto
a halo of any mass (black curves) or onto haloes more massive than $2.6\times10^{9}\Mass$.
\label{fig:redshift_fo}} 
\end{figure}

\begin{figure} 
\includegraphics[width=8.5cm]{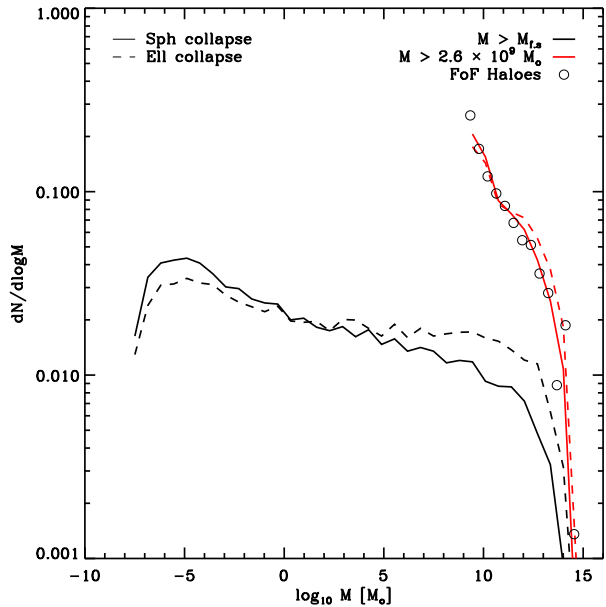} 
\caption{ The mass distribution at first threshold crossing.
The solid and dashed lines indicate the results for constant and moving
collapse barriers, respectively.  Note that the statistics only included the
trajectories that do eventually fulfill the criterion for collapse, hence, the
area below each curve is 1. 
\label{fig:mass_fo}} 
\end{figure}

In Fig.\ref{fig:redshift_fo} we present the distribution of the redshift at
which the threshold is first crossed for the random walks that end up in some
halo at $z=0$. This corresponds to the differential probability that a halo
particle first becomes part of a collapsed object at redshift $z$.

If we assume the ellipsoidal model for collapse (the dashed lines), the median
value of the distribution is $z=10.65$, and $90\%$ of the crossings occur at
redshifts lower than $35.2$. For the spherical collapse (solid lines) these
redshifts are modified to $z=12.8$ and $z=33.8$ respectively. Thus less than
half of the neutralinos that are part of DM haloes today were part of any halo
at $z=13$. Less than $10\%$ of the mass in today's haloes was already in a halo
of any mass by redshift $36$. 

Fig.\ref{fig:mass_fo} shows the distribution of the smoothing masses at which
these first barrier crossings occur. This plot thus displays the probability
per logarithmic mass interval that a neutralino that is part of a halo today
was first accreted in diffuse form onto a halo of mass $M$.  This is equivalent
to the mass distribution of the first haloes of randomly chosen DM particles
from present-day haloes.

Because of the shallow slope of the variance at small radii (Fig. 2), the
distribution of masses shown in Fig.\ref{fig:mass_fo} is extremely flat; it
varies by a factor of $\sim3$ across $\sim20$ orders of magnitude in mass. As a
result, the first halo mass for DM particles is very diverse. Median values of
the distribution are $\sim 10^{-2}\Mass$ and $1.4\Mass$ for the constant and
moving barrier respectively.  Most neutralinos were never part of a halo of
mass $M < 10^5\,M_{f.s.}$. In fact, only $10\%$ were ever part of an object
smaller than Earth mass, the same mass fraction for which the first halo was
more massive than $\sim 10^7\Mass$.

Further results from our random walks are that the median redshift of collapse
of ``first" objects of mass smaller than $10^{-4}\Mass$ is $24.1$, only
slightly larger than that of objects in the range $10^{-2}\Mass$ to
$10^{2}\Mass$ which is $z = 15.8$.  This suggest that the concentration of the
smallest objects in the CDM hierarchy should not be significantly greater than
that of objects thousands of times more massive. Thus we may expect that most
objects with mass $M\sim M_{f.s.} $ are efficiently disrupted by tidal forces
as they merge into more massive haloes.

In both plots the measurements from the hMS simulation are compared with
predictions from the excursion set theory, where we impose an artificial mass
cut to match the resolution of the N-body simulation (red lines). The simulated
redshift and mass distributions agree well with those predicted by the
excursion set formalism. This motivates our extrapolation of the predictions
down to scales where the theory has not yet been tested.  Nevertheless, it is
important to keep in mind that this is an aggressive extrapolation.

\section{The assembly of haloes}

The particle-based nature of the excursion set formalism makes it ideal to
study the assembly history of dark matter haloes. As we travel along a random
walk we follow a DM particle from the very bottom of the CDM hierarchy through
the successive mergers it witnesses until it ends up, for example, in a
Milky-Way mass halo today. For the statistics in this section we have modified
our ensemble of trajectories so that all of them end up in a halo of similar
mass to the one that hosts our own galaxy ($M = 1.37 \times10^{12}\Mass$).

\subsection{Infall events}

\begin{figure} \includegraphics[width=8.5cm]{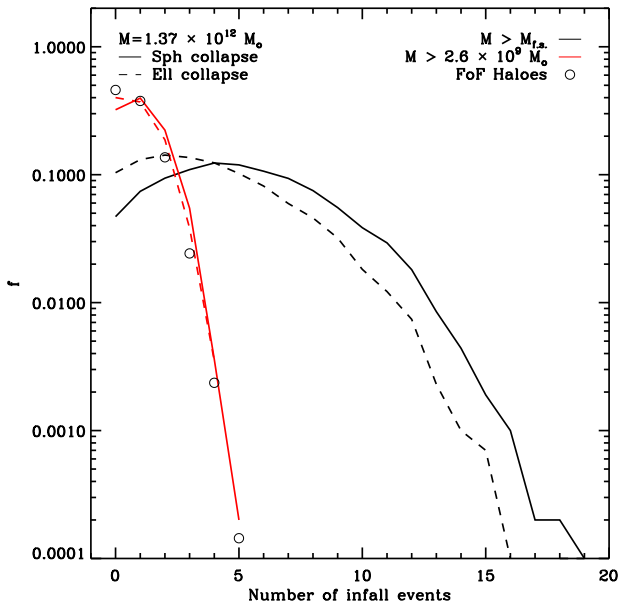} 
\caption{ A histogram of the number of infall events for a random sample of all the
dark matter particles that constitute a $1.37 \times 10^{12}\Mass$ halo today. 
Sets of curves indicate the number of infalls since the DM particles were accreted
onto any halo (black lines) or onto a $2.6\times10^9\Mass$ halo (red lines). Solid and dashed
lines assume a constant and moving barrier, respectively. 
\label{fig:ninfall}} \end{figure}

Fig.\ref{fig:ninfall} presents the distribution of the number of infall events
for all the trajectories in our ensemble of random walks. These are defined as
an increase in mass by more than a factor of two, and so are the events where
the halo that hosts a given particle is accreted onto a larger system. 

Black lines in Fig.\ref{fig:ninfall} follow DM particles over the full
hierarchy, while red lines follow them only from the moment they are hosted by
a halo more massive than $2.6 \times 10^{9}\Mass$. The latter results can be
compared directly to our N-body simulation.  As in previous plots, results for
both the spherical collapse barrier (solid lines) and the square root barrier
(dashed lines) are shown in the figure.

If most mergers were to occur between objects of similar mass, the typical
number of infall events would be very large. The difference in mass between the
smallest structures and a Milky-Way halo is $\simeq 20$ orders of magnitude, so
one might expect to see typically $20\,\log_{2} 10 \sim 67$ infall events for
each particle. Fig.\ref{fig:ninfall} indicates that this is a poor
representation of structure assembly in a CDM Universe.  The median of the
histogram is between 4 and 5 events. Thus, a typical mass element in a
Milky-Way halo has gone through a rather small number of events where its host
halo falls into something bigger than itself. Among all these infall episodes,
typically only one occurs while the particle is part of a halo larger than $2.7
\times 10^9\Mass$.  This result suggests that minor mergers play an important
role in the growth of dark matter haloes. 

Another common misconception is that the growth of dark matter haloes is due
entirely to the accretion of other haloes. From our random walks we can assess
this statement by computing the amount of mass that is formally accreted
smoothly, i.e. not via mergers of any type. This quantity can be read off from
Fig.\ref{fig:ninfall} as the fraction of mass that has suffered zero infalls.
We can see that, depending of the shape of the barrier, $5-10\%$ of the mass of
a $\sim 10^{12} \Mass$ halo was accreted smoothly. 

Naturally by imposing a higher mass threshold the amount of unresolved
accretion increases, reaching $\sim 40\%$ when the minimum mass corresponds to
the resolution limit of our N-body simulation. This matches the values we
measure directly. 

The results presented in this subsection are weakly dependent on the mass of
the final halo.  Even small haloes, of about one solar mass, have accreted
$10-20\%$ of their mass smoothly and their particles have typically suffered
$1-2$ infall events.

\subsection{The distribution of mergers}

\begin{figure} 
\includegraphics[width=8.5cm]{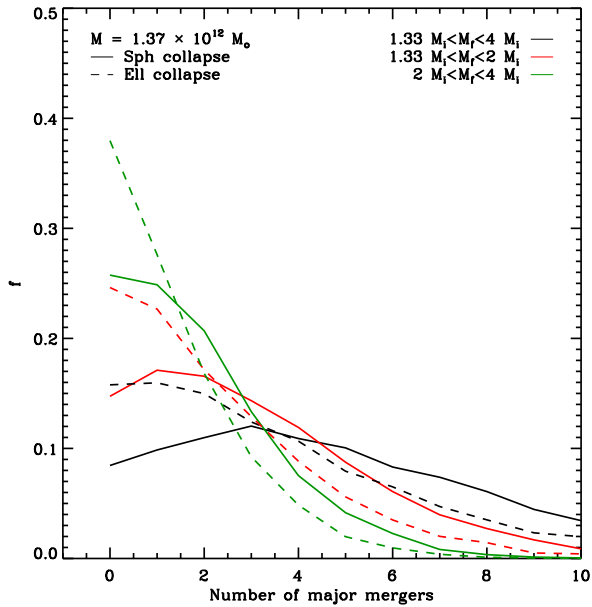} 
\caption{ A histogram of the number of major mergers in which the dark matter particles
of a Milky-Way sized halo have been involved throughout their history. Black curves 
histogram the total number of major mergers experienced by the haloes containing 
given dark matter particles while red curves include only those major mergers
where the particle is part of the larger object and green curves those where it is
part of the smaller object. 
\label{fig:nmergers}}
\end{figure}

To close this section, in Fig.\ref{fig:nmergers}  we turn our attention to the
number of major mergers that a typical DM particle that ends up in a Milky-Way
sized halo has experienced during its life. We define a major merger as an
event in which the halo associated to a given trajectory changes its mass by a
factor in the range [1.33, 4]. This represents mergers where the mass ratio of
the haloes is less than a factor of three. 

The histograms displayed in Fig.\ref{fig:nmergers} show that most dark matter
particles have been involved in a relatively small number of major mergers.  In
fact, $10-20\%$ of all particles never witness a major merger.  The median of
the distribution is $4$ for the spherical collapse model and $3$ for the square
root barrier.

We divide the trajectories into two categories according to the size of the
halo before the merger.  The first subgroup, displayed as red lines, represents
the cases where the particle belongs to the larger object ($1.33 M_i < M_f < 2
M_i$) while green lines show the case where the particle belongs to the smaller
object ($2 M_i < M_f < 4 M_i$). It is interesting to note that these subsets
have different distributions. This implies that major mergers where the
particle is part of the smaller object are more frequent than those where it is
part of the larger one.

The mass cut-off induced by free streaming also produces a correlation between
the redshift and the mass ratio of the progenitors. Mergers at high redshift
are more likely to have low mass ratios, while mergers between very dissimilar
haloes necessarily occur only at low redshift.  

\section{Conclusions}

We have used the excursion set formalism to examine halo formation and mass
assembly in a standard $\Lambda$CDM cosmology where the DM is made of $100$ GeV
neutralinos. Our analytic treatment of the problem allows us to follow halo
assembly histories for a large number of DM particles over more than 25 orders
of magnitude in halo mass. For the first time we are able to study halo
assembly histories of DM particles from the very bottom of the CDM hierarchy up
to the largest structures that exist today.  

The free streaming of neutralinos induces an exponential cut-off at large
wavenumbers in the primordial power spectrum. For our neutralino model this
sets $10^{-8}\Mass$ as the minimum mass that any dark matter halo can have. Our
results suggest, however, that a very small fraction of the DM particles were
ever part of a halo of mass $\sim M_{f.s.}$. In fact, the mass of the first
halo for a typical particle is comparable to the mass of the Sun. In addition,
we have found that most of the matter is not part of any halo at early times;
the typical redshift for first collapse is $z=14$.  Today, $5\%$ to $10\%$ of
the dark matter is still not part of any clump, and beyond redshift $14$ most
of the mass was in diffuse form. These late formation times imply that even
very low mass haloes are not expected to be strongly concentrated and thus that
they should be relatively easily disrupted.

The particle-based formulation of the excursion set theory allowed us to trace
back to the very bottom of the CDM hierarchy, the particles that today form a
Milky-Way sized halo. We found that there are rather few generations of
accretion/merger events. Typical particles experience three or five of such
episodes, only one of which occurs after the particle is part of a
$10^{9}\Mass$ halo.  About $10\%$ of the mass of Milky-Way sized haloes was
accreted in diffuse form rather than as part of a smaller halo.

Our results depend little on the exact mass of the neutralino or on the shape
of the barrier for collapse. Comparison with an N-body simulation suggests that
the excursion set formalism gives reliable results at least over the limited
mass range where a comparison can be made. Structure growth in the concordance
cosmology is considerably less hierarchical than is often thought. 

\section*{Acknowledgments}

We would like to thank Adrian Jenkins for providing us with the N-body
simulation used in this work. We also acknowledge useful conversations with
Cedric Lacey, Chung Pei Ma, Jorge Moreno and Eyal Neistein.

\bibliographystyle{mn2e} \bibliography{eps}

\label{lastpage} \end{document}